\documentclass[
 reprint,
 amsmath,amssymb,
 aps,
pra,
noeprint
]{revtex4-2}

\usepackage{graphicx}% Include figure files
\usepackage{dcolumn}% Align table columns on decimal point
\usepackage{bm}% bold math
\begin{document}
	\title{Optimization of experimental parameters for laser-slowing and magneto-optical trapping of MgF molecules}
	\author{Dongkyu Lim}
    \affiliation{Department of Physics, Korea university, Seoul 02841, Republic of Korea}

    \author{Eunmi Chae}
    \email{echae@korea.ac.kr}
    \affiliation{Department of Physics, Korea university, Seoul 02841, Republic of Korea}
    
    \begin{abstract}
        Diatomic molecules are promising systems for quantum science applications due to their complex energy structures and strong dipole-dipole interactions. 
        Achieving ultracold temperatures is essential for these applications, but the complexity of molecular energy levels requires precise optimization of experimental parameters for laser slowing and magneto-optical trapping (MOT). 
        Here, we simulate and optimize the complete process of slowing and trapping MgF molecules, from a buffer-gas beam source to MOT capture, using Bayesian optimization. 
        By combining laser slowing and MOT simulations, we identify parameters that maximize the capture velocity and the ratio of trapped molecules. 
        Our results demonstrate a maximum MOT capture velocity of 82.5 m/s, and 28.6\% of the molecules that reach the MOT region are trapped under optimal conditions. 
        These findings provide insights into experimental setups for MgF and similar molecules, offering a framework for advancing molecular laser cooling and quantum experiments.

    \end{abstract}
    \maketitle

\section{Introduction}\label{sec:introduction}
	Diatomic molecules have been proposed as an attractive system for quantum simulation/computation\cite{2002Demille,2009Carr,2012Baranov,2018KangKuenNi,blackmore2018ultracold,Yan2013,hughes2020robust,sawant2020ultracold,cornish2024quantum,Chae2024}, quantum chemistry\cite{bohn2017cold, cheuk2020observation,balakrishnan2016perspective, liu2020photo, hu2021nuclear,hu2019direct,doyle2021science} and precision measurements\cite{hudson2011improved, acme2018improved} with their abundant internal energy structures and strong electric dipole-dipole interactions.  
    Ultracold temperatures are prerequisite to maximize the merits of the molecules.  
    Laser cooling and a magneto-optical trap (MOT) are the most frequently used techniques to reach ultracold temperatures for atoms.
    To expand these techniques to molecules, their complex energy structures must be addressed adequately \cite{Demille2010laser,JYe2008magneto,chae2023laser,FITCH2021157}. 
    This results in utilizing multiple lasers, and therefore one needs to optimize many parameters during the experiment including frequencies, powers, and polarizations of the lasers. 
    
    The slowing and trapping processes for diatomic molecules using lasers are more challenging than for typical alkaline atoms due to the presence of ro-vibrational states and complex hyperfine structures.
    Since vibrational transitions have no selection rule, multiple lasers are required to achieve a quasi-closed cycling transition.
    Moreover, Zeeman slowing, an efficient laser-slowing technique used for atoms, becomes challenging to implement to molecules due to molecules' complex energy structure.
    Thus, a different slowing procedure is necessary, such as white-light and frequency-chirping.
    Additionally, to slow down a molecular beam from a buffer-gas beam source, the experimental parameters of the slowing laser should be carefully selected based on the initial velocity distribution of the molecules.
    Therefore, optimizing experimental variables one by one would be a time-consuming task in real experiments.

    Simulations can help optimize multiple parameters of the experiment. 
    The interactions between the molecules and the environment including lasers and magnetic fields are well known and easily implementable to simulations.
    With properly chosen rewards, such as the number of trapped molecules inside the MOT, all experimental                                     parameters can be optimized using well-established optimization algorithms. 
    Numerical simulations for diatomic molecules have predominantly been conducted to optimize the MOT, determining the trapping force and equilibrium temperature\cite{tarbutt2015magneto,xu2019three} or identifying the capture velocity\cite{xu2021maximizing}.
    However, simulations of the slowing stage are another important part in reducing the velocities of molecules enough to be captured in the MOT\cite{yan2022simulation}.
    Both the slowing stage and the MOT stage should be considered together in the simulation to determine the experimental parameters that are suitable for real experiments.
    Moreover, details of molecular beams, such as forward velocity and transverse divergence, as well as different Zeeman shifts of molecules at various regions, are all important for the simulation. 
    Therefore, a combined simulation of slowing and MOT is necessary to acquire relevant insights about the molecular experiment.
    
	In this paper, we find the best experimental parameters for the MOT and slowing to capture the maximum number of MgF molecules in a MOT from a buffer-gas beam source using Bayesian optimization. 
    MgF is expected to have a higher number of molecules captured in a MOT than other alkaline earth metal–ligand diatomic molecules due to its relatively low mass and short main transition wavelength of 359 nm, which provides a large momentum kick.
    We simulate the motions of the molecules as they decelerate from the buffer-gas beam source to the MOT and find the expected ratio of the trapped molecules to the total molecules that reach the MOT chamber. 

\section{Background Theory}\label{sec:background theory}
	\subsection{Energy structure of MgF molecules}\label{subsec:energy structure of MgF molecules}

    \begin{figure}[h]
        \centering
        \includegraphics[width=8.5cm]{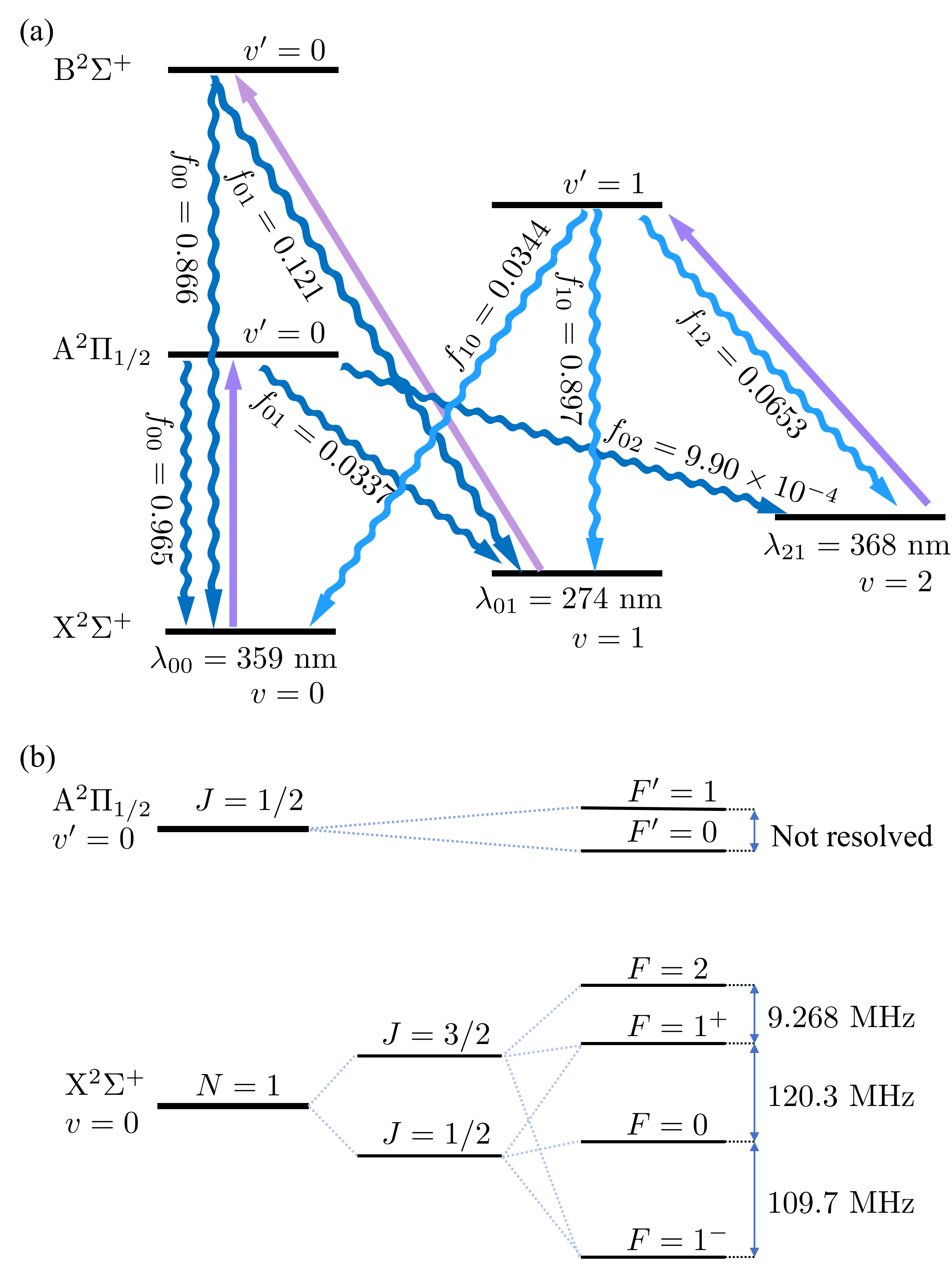}
        \caption{MgF energy structure for (a) electro-vibrational states and (b) hyperfine states. 
                    (a) Main transition line is from $|\mathrm{X}^2\Sigma, v=0\rangle \rightarrow |\mathrm{A}^2\Pi_{1/2}, v'=0\rangle$ at 359 nm.
                    $f_{nn}$ indicates the Franck-Condon factor(FCF) between two vibrational states.
                    Two repump lasers($|X^2\Sigma,v=1\rangle\xrightarrow{} |B^2\Sigma,v'=0\rangle$ and $|X^2\Sigma, v=2\rangle \xrightarrow{} |A^2\Pi_{1/2},v'=1\rangle$) are required to increase the number of photon scattering up to $2\times 10^4$\cite{norrgard2023radiative}. 
                    (b) Energy difference between $F'=0$ and $F'=1$ in $\textit{A}^2\Pi_{1/2}$ state are not resolved yet.
                    Note that splitting between $F = 2$ and $F=1^+$ in $\textit{X}^2\Sigma^+$ state is smaller than $0.5\Gamma$.}
        \label{fig:MgF energy structure}
    \end{figure}

	MgF is one of the alkaline-earth monofluoride molecules, same as SrF and CaF that have been trapped in MOTs using their quasi-closed cycling transitions.
    MgF has a light mass and a main transition wavelength of $359$ nm, which can provide a significant deceleration to the molecule during photon scattering.
    As a result, MgF is expected to have a higher efficiency for laser-cooling compared to other heavier alkaline-earth monofluoride molecules.
	
	Figure \ref{fig:MgF energy structure} shows the hyperfine splitting of MgF molecule.
    $X^2\Sigma^+ (v=0,\ N=1,\ -)$ state and $A^2\Pi_{1/2} \left(v'=0,\ J'={1}/{2},\ +\right)$ state are chosen as the ground and the excited states respectively for optical cycling, where $v$ is the vibrational quantum number, $N$ is the rotational quantum number, $J,\ J'$ is the total angular momentum quantum number without nuclear spin, and ($\pm$) denotes parity.
    These two states can form quasi-closed cycling transitions due to selection rules.
    The excited state with $J' = {1}/{2}$ can only decay to $J = {1}/{2}$ or $J={3}/{2}$ because the electric dipole (E1) transition must change the parity and allow the change of $J$ only for $\Delta J = 0,\pm 1$. 
    The ground state has four hyperfine states, $|F = 1^-,0,1^+,2\rangle$, while the excited state has two hyperfine states, $|F' = 0,1\rangle$, where $F$ and $F'$ are the total angular momentum quantum numbers including nuclear spin.
    Note that the energy gap between the hyperfine structure in \textit{A} state has not been resolved yet.
    The natural linewidth of the $A^2\Pi_{1/2}$ state is $\Gamma =2 \pi \times 20.9$ MHz, corresponding to a lifetime of $\tau = 7.60$ ns \cite{norrgard2023radiative}.
    Since the energy splitting between $|F = 1^+\rangle$ and $|F=2\rangle$ is about half of the natural linewidth, we can use the lasers with either three or four frequencies to cover the hyperfine states.
    The g-factors of $X^2\Sigma^+$ state $|F = 2,1^+,1^-\rangle$ are $g_2\approx0.5$, $g_{1^+}\approx0.71$ and $g_{1^-}\approx-0.21$\cite{xu2019three}.
    For $A^2\Pi_{1/2}$ state, g-factor of $|F' = 1\rangle$ is estimated to be $g_e\approx-0.0002$\cite{xu2019three}, which is much smaller than CaF ($g_e \approx -0.021$) \cite{tarbutt2015modeling}.
	
    \subsection{Calculation methods}
	
    In this work, we utilize rate-equation modeling combined with Bayesian optimization to identify the optimal parameters for molecular cooling and trapping through motion simulations.
    This model evaluates the optical force by calculating the net momentum transfer resulting from photon-scattering processes, thus establishing a connection between the dynamics of the internal-state population and the translational motion of the molecule\cite{tarbutt2015magneto}.
    Although this approach neglects coherent effects, its significantly lower computational cost enables rapid simulation of molecular motions.
    These calculations are performed using PyLCP, a Python package developed for computing the laser cooling in AMO physics with various experimental conditions\cite{eckel2022pylcp}.
    
    Bayesian optimization is used to determine the optimal set of experimental parameters in the simulation.
    Evaluating and optimizing outputs, such as the capture velocity or the trapped ratio, involves a complex process.
    Bayesian optimization is appropriate in this setting because it does not require derivatives of the objective and is effective in high-dimensional spaces\cite{jones1998efficient, shahriari2015taking,frazier2018tutorial}. 
    We employ the Bayesian Optimization Python package to perform this optimization in conjunction with our PyLCP-based simulations\cite{Fernado2014Bayesian}.

    % The simulation involves optimizing numerous experimental parameters to maximize outputs such as the capture velocity or the trapped ratio of molecules. 
    % Since evaluating this objective function is computationally expensive and exists within a high-dimensional search space, Bayesian optimization is a well-suited method for this task. 
    % To implement this, we use the Bayesian-optimization package in Python, which is compatible with our PyLCP-based simulation environment\cite{Fernado2014Bayesian}.
    
	\section{Simulation process}\label{sec:simulation process}
	
	\subsection{Design of the experimental setup}\label{subsec:design of the experimental setup}
	\begin{figure*}[t]
		\centering
		\includegraphics[width=17.8cm]{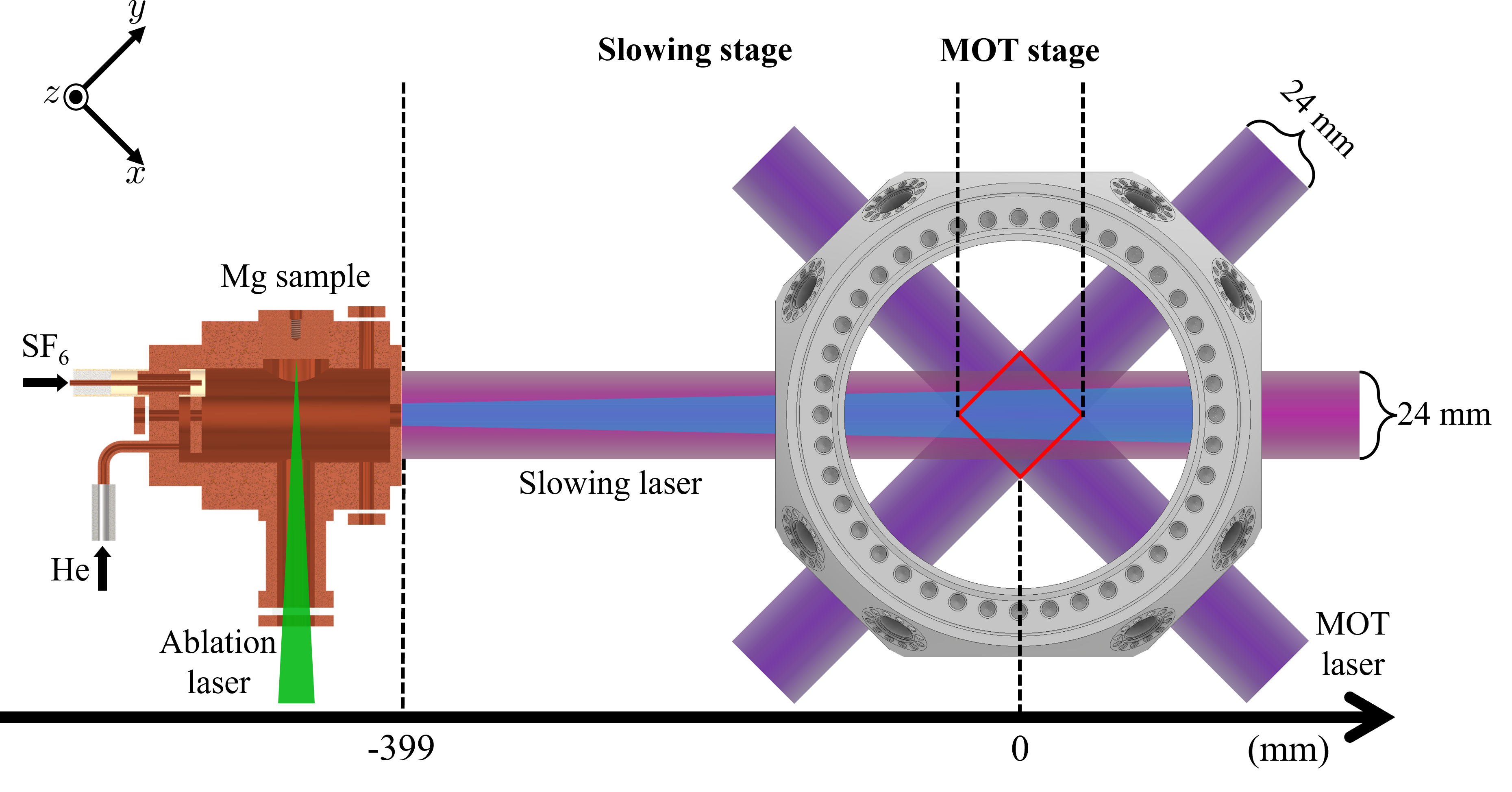}
		\caption{Schematic of the experimental setup for the simulation. 
        After creation by ablation, MgF molecules are cooled down by He buffer-gas and emerge out from the buffer-gas cell as a MgF molecular beam (blue line), heading to an octagonal MOT chamber. 
        The MOT stage is the region where six laser beams are crossed (inside the red square), and the slowing stage is the region from the exit of the buffer-gas cell to the entrance of MOT stage.
        Slowing lasers and MOT lasers are indicated as purple lines.
        Magnetic field is applied by anti-Helmholtz coils along the $z$ axis.
        }
		\label{fig:Scheme}
	\end{figure*}
 
	The design of simulation process is shown in Fig.\ref{fig:Scheme}.
    MgF molecules are generated and cooled in the He buffer-gas cell and travel in the $xy$ plane at a ${\pi}/{4}$ angle from the $x$ axis, heading toward the center of the MOT chamber. 
    The molecular beam emerging from the buffer-gas cell is assumed to have an initial longitudinal velocity distribution with a mean of 107 m/s and a full width half maximum (FWHM) of 74.4 m/s, and the transverse velocity has the same FWHM as 74.4 m/s while the mean value is zero. 

    The overall simulation process consists of two stages : the MOT stage and the slowing stage.
    In the MOT stage, six-way lasers along the three-axis and a magnetic field from anti-Helmholtz coils are applied to trap the MgF molecules in a MOT.
    The MOT laser has four frequency components to cover the four hyperfine structure components of the $X^2\Sigma^+$ state.
    The laser, consisting of four frequencies, is equally divided into three directions and retro-reflected along each axis, resulting in a total of six beams directed toward the center of the MOT chamber.
    The total power of the laser is 600 mW, divided into 200 mW for each three-axis.
    Once the motions of the MgF molecules inside a MOT are calculated with the given initial conditions, we determine the capture velocity by checking their motional trajectories.
    By maximizing the capture velocity, the optimal experimental parameters can be found for the MOT stage .
    
    In the slowing stage, the slowing laser is turned on for a certain amount of time in order to decelerate the MgF molecules enough to be trapped in a MOT. 
    The slowing laser passes through two electro-optic modulators (EOM).
    One EOM with the modulation frequency of approximately 120 MHz covers the hyperfine structure. The other EOM for white-light slowing has a modulation frequency of about 20 MHz to cover the Doppler shift as the molecules slow down.
    The modulation frequencies are set as parameters of the optimization. 
    The total laser power is 600 mW, which is a combination of two 300 mW laser beams with $\sigma^+ $ and $ \sigma^-$ polarization.
    In the slowing stage, the ratio of the trapped molecules in a MOT to the total molecules that reach the MOT chamber is calculated.
    By maximizing this ratio, the optimal experimental parameters for the slowing stage can be determined.
    In general, we first simulate the MOT stage and then simulate the slowing stage subsequently using the optimized experimental condition in the MOT stage.
	
	\subsection{MOT stage}\label{subsec:MOT stage}
	
	% \subsubsection{Parameter set}
	\subsubsection{Model setup}\label{subsubsec:MOT model setup}

    \begin{table}[t]
    \caption{Parameter ranges used in the MOT-stage optimization.
    Four laser beams whose index are labeled as $i=1,2,3,4$ cover the transitions from $|F = 1^-,0,1^+,2\rangle$ to excited state $|J'=1/2\rangle$.
    Each laser has different frequency detuning $\left(\delta_i\right)$, relative power $\left(P_i\right)$ and polarization $\left(\sigma_i\right)$.
    Frequency detuning $\delta_i$ is expressed in units of the natural linewidth $\Gamma$.    
    Relative power $P_i$ refers to the ratio of the power of each frequency component to the total laser power.
    Polarization $\sigma_i$ contains the entire configuration of six-way laser components used in a MOT. $(+)$ means that the polarization of the laser along the $z$ axis is $\sigma^+$ and $\sigma^-$ for the ${x},{y}$ axes. 
    $(-)$ has the opposite configuration with $(+)$.
    Magnetic field is applied as $\Vec{B} = \alpha (-x/2,-y/2,z)$, where $\alpha$ is given in G/mm.}
    \label{table:MOT range}
    \begin{ruledtabular}
    \begin{tabular}{lcc}
        Parameter & Symbol & Range \\
        \hline
        Frequency detuning & $\delta_i$ & $[-5, 5]$ \\
        Relative power & $P_i$ & $[0, 1]$ \\
        Polarization & $\sigma_i$ & $(+)$ or $(-)$ \\
        Magnetic field gradient & $\alpha$ & $[0, 2]$ \\
    \end{tabular}
    \end{ruledtabular}
    \end{table}
	
	The parameters to be optimized and their set ranges are listed in Table. \ref{table:MOT range}.
    The total 13 experimental parameters are combined into a parameter vector $\vec{X}_{\rm MOT} = (\delta_i, P_i, \sigma_i,\alpha)$. 
    For a given $\vec{X}_{\rm MOT}$, we simulate the motion of MgF molecule with various initial velocities $v_l$ using the rate equation.
    In an initial step, the transverse positions and the transverse velocities are set to zero.
    The molecules are determined to be "trapped" when their final positions are within 6 mm in each axis, with velocities below 0.7 m/s directed towards the center, where the temperature of the MgF molecule is below 1 mK.
    Otherwise, the molecules are determined to be "lost".
    The maximum initial velocity that the molecules are trapped is defined as the capture velocity $v_c$ for the given experimental parameters.
    Bayesian optimizing model searches for $\Vec{X}_{\rm MOT}$ that yields the largest $v_c$.
    This optimal $\Vec{X}_{\rm MOT}$ will later be used to find the ``MOT-trappable condition function", which considers the initial transverse positions and velocities of the molecules.
	
	\subsubsection{Result}\label{subsubsec:MOT result}
    
    \begin{table*}[t]
    \centering
    \caption{Optimal parameters with maximum $v_c = 82.5$ m/s.
    Relative powers are expressed as percentages, with the total set to 100$\%$.}
    \label{table:MOT result}
    \begin{ruledtabular}
    \begin{tabular}{lccccccccccccc}
    Case & $\alpha$ & $\delta_1$ & $\delta_2$ & $\delta_3$ & $\delta_4$ 
         & $P_1$ & $P_2$& $P_3$ & $P_4$ 
         & $\sigma_1$ & $\sigma_2$ & $\sigma_3$ & $\sigma_4$ \\
    \hline
    1  & 0.628 & -2.99 & -0.923 & -0.079 & -0.541 & 14.7 & 24.6 & 33.6 & 27.1 & $(+)$ & $(-)$ & $(+)$ & $(-)$ \\
    2  & 0.594 & -3.11 & -1.268 & -0.274 & -0.483 & 14.0 & 25.8 & 32.7 & 27.5 & $(+)$ & $(-)$ & $(+)$ & $(-)$ \\
    3  & 0.589 & -3.06 & -1.287 & -0.043 & -0.334 & 11.7 & 28.3 & 30.0 & 30.0 & $(+)$ & $(-)$ & $(+)$ & $(-)$ \\
    4  & 0.528 & -3.03 & -1.391 &  0.079 & -0.479 & 13.9 & 16.0 & 42.1 & 28.0 & $(+)$ & $(-)$ & $(+)$ & $(-)$ \\
    5  & 0.620 & -3.00 & -1.217 & -0.195 & -0.303 & 12.5 & 27.7 & 38.7 & 21.0 & $(+)$ & $(-)$ & $(+)$ & $(-)$ \\
    6  & 0.817 & -3.07 & -1.230 & -0.215 & -0.558 & 16.9 & 20.5 & 33.7 & 28.8 & $(+)$ & $(-)$ & $(+)$ & $(-)$ \\
    7  & 0.819 & -2.82 & -1.240 & -0.236 & -0.728 & 19.3 & 12.7 & 52.4 & 15.7 & $(+)$ & $(-)$ & $(+)$ & $(-)$ \\
    8  & 0.857 & -3.14 & -1.649 & -0.327 & -0.771 & 23.9 &  9.4 & 51.3 & 15.5 & $(+)$ & $(-)$ & $(+)$ & $(-)$ \\
    9  & 0.491 & -2.94 & -1.524 & -0.156 & -0.682 & 22.7 & 16.1 & 47.2 & 14.0 & $(+)$ & $(-)$ & $(+)$ & $(-)$ \\
    10 & 0.737 & -3.06 & -1.073 & -0.374 & -0.685 & 21.4 & 12.7 & 45.2 & 20.6 & $(+)$ & $(-)$ & $(+)$ & $(-)$ \\
    11 & 0.711 & -2.87 & -1.149 & -0.267 & -0.414 & 20.1 & 12.9 & 41.7 & 25.3 & $(+)$ & $(-)$ & $(+)$ & $(-)$ \\
    12 & 0.494 & -2.85 & -1.052 & -0.187 & -0.594 & 23.5 & 16.0 & 40.0 & 20.4 & $(+)$ & $(-)$ & $(+)$ & $(-)$ \\
    Mean & 0.720 & -3.00 & -1.174 & -0.187 & -0.568 & 17.8 & 19.7 & 39.9 & 22.6 & $(+)$ & $(-)$ & $(+)$ & $(-)$ \\
    \end{tabular}
    \end{ruledtabular}
    \end{table*}

	Table \ref{table:MOT result} shows twelve optimized results with a maximum capture velocity of $v_c = 11\ \Gamma/k \simeq 82\ \rm m/s$, and the mean value of the set of parameters. 
	In all the cases, all the parameter values are similar to each other.
    For frequencies, the red detuning of $\delta_1$ is the largest about $3.0\ \Gamma\simeq 63$ MHz.
    The power ratio of laser 3 is larger than the others.
    Although the relative magnitudes of the lasers are changed, the ratio between $P_1 +P_2$ and $P_3 + P_4$ is constant about $4:6$.
    All polarizations are fixed as $\left(\sigma_1, \sigma_2, \sigma_3, \sigma_4\right) = (+,-,+,-)$, and the magnetic field gradient $\alpha$ is also in the range between $(0.4,0.9)$.
    Since the optimal parameter values for twelve cases have a similar aspect, we can estimate that the mean value of each parameter $\vec{X}_{\rm mean}$ can also provide the optimized result.

    \begin{figure}
        \centering
        \includegraphics[width=8.5cm]{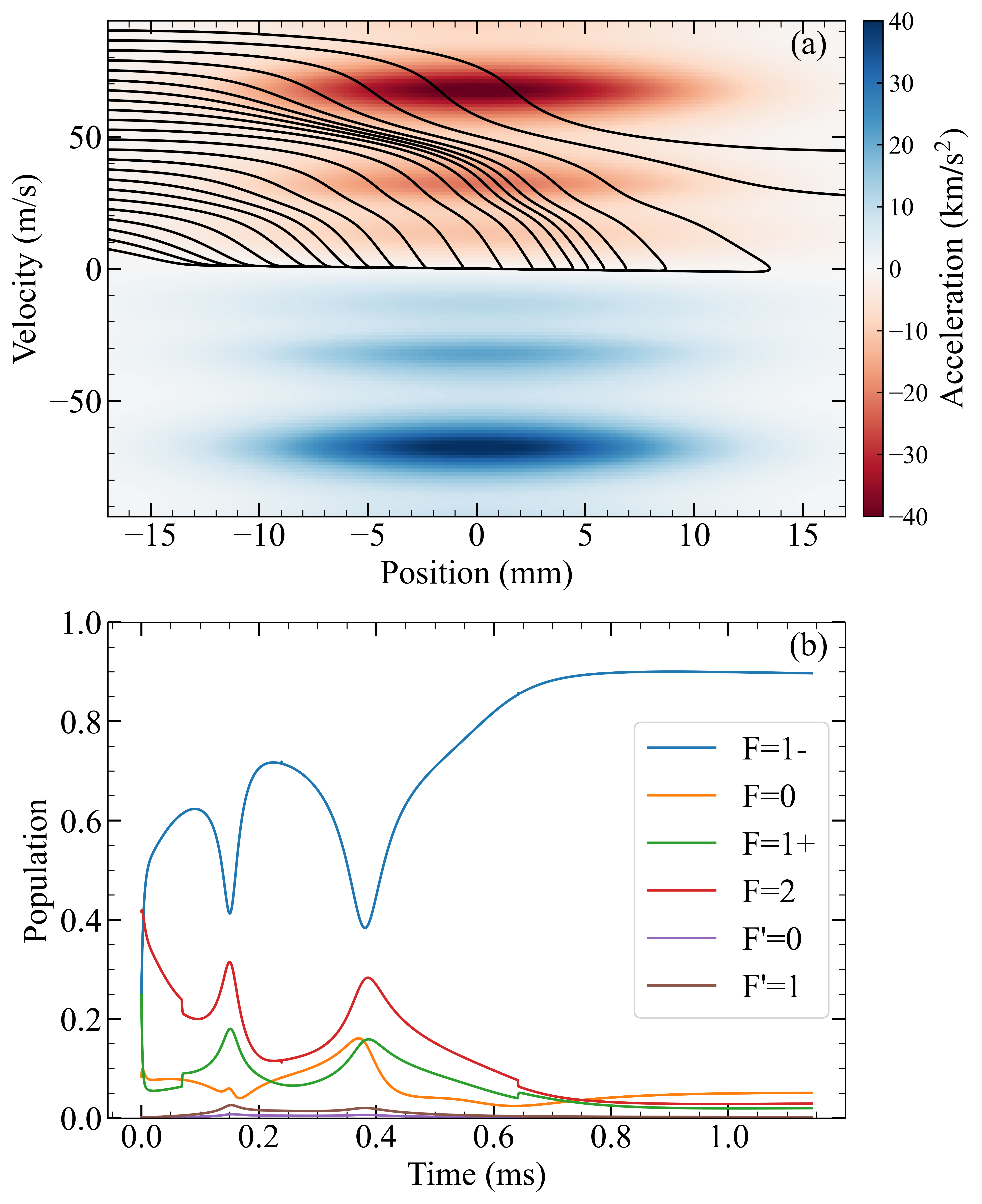}
        \caption{(a) Acceleration on the MgF molecules and their calculated trajectories for the mean parameter set. Longitudinal axis has an angle of ${\pi}/{4}$ from the $x$ axis in the $xy$ plane. Acceleration is shown with red or blue color. 
        Each trajectory line has different initial velocity from 7.5 m/s to 90 m/s with 3.75 m/s intervals. 
        (b) State population vs time of molecules traveling in a MOT chamber with initial longitudinal velocity 82.5 m/s. 
        Six colors indicate the sum of the populations in all states with the same $F$ number.}
        \label{fig:MOT_final_result_avg}
    \end{figure}

    Fig.\ref{fig:MOT_final_result_avg} (a) shows the acceleration and trajectories of the molecules with the parameter $\vec{X}_{\rm mean}$. 
    The molecules with initial longitudinal velocity from 7.5 m/s to 82.5 m/s converge well to the origin, which means that the molecules are well trapped in a MOT. 
    This indicates that $\vec{X}_{\rm mean}$ is also the case for optimal capture velocity. 
    The darkest color region, where the molecules are rapidly decelerated, exists between $50\sim75$ m/s region.
    Fig.\ref{fig:MOT_final_result_avg} (b) shows the change of population in each state of the molecules during the motion in a MOT. 
    The ratio of the molecules in the $|F=1^-\rangle$ state is dominant, over $90$ $\%$ after 1 ms. 
    This is because $\delta_1$ is larger and $P_1$ is smaller than the others, so the molecules remain in $|F=1^-\rangle$ mostly as the velocity and position of the molecules converge to zero. 
    There are two wells(hills) of $|F=1^-\rangle$(the other ground states) in Fig.\ref{fig:MOT_final_result_avg} (b), which indicates that there are more photon scattering events, thus the molecules decelerate a lot at these points.

    The significant accumulation of the population in the $|F=1^-\rangle$ state indicates that $|F=1^-\rangle$ plays a crucial role in determining the optimal parameter set.
    The relative transition strengths of $|F=1^+\rangle$ and $|F=1^-\rangle$ are 0.422 and 0.411, which are the largest among all transitions from the $X^2\Sigma^+$ state.
    Although the transition strength for $|F=1^+\rangle$ is higher than that for $|F=1^-\rangle$, the close proximity of $|F=1^+\rangle$ and $|F=2\rangle$ implies that the $|F=1^+\rangle$ state cannot be used selectively.
    Consequently, in the laser cooling process, the $|F=1^-\rangle$ guarantees the highest photon scattering rate, so accumulating the population toward $|F=1^-\rangle$ is advantageous for achieving higher capture velocities.
    The task of directing the population towards $|F=1^-\rangle$ is primarily facilitated by the $z$ axis beam. This is evident from the larger frequency detuning and low power of laser 1, which is designed to pump the population from other states into $|F=1^-\rangle$ through optical pumping.

    The main cooling process for the molecules in $|F=1^-\rangle$ is facilitated by the beams along the $x$ and $y$ axes. 
    Suppose that molecules entering the MOT stage initially have a velocity of $82.5$ m/s.
    Because of Doppler effect, the molecules would see all laser frequencies as being blue detuned by approximately $11.0\ \Gamma/\sqrt{2}\simeq163$ MHz.
    The frequencies of lasers 3 and 4 with about $63$ \% of the total power are initially resonant with the fast molecules in the $|F=1^-\rangle$ state, leading to a large scattering rate and effective deceleration.

    \begin{figure}[h]
        \centering
        \includegraphics[width=8.0cm]{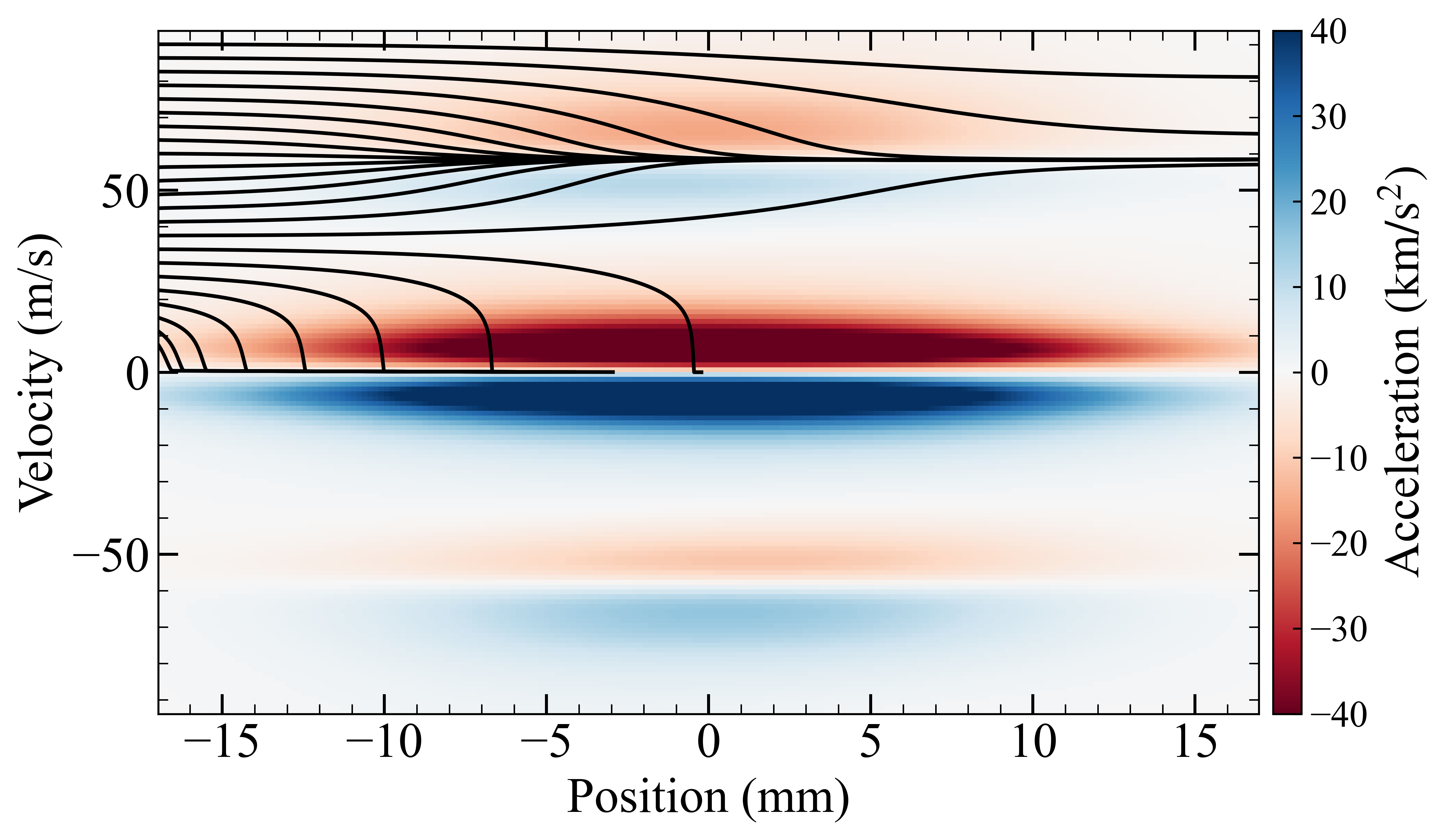}
        \caption{Acceleration and motional trajectories of the molecules with a conventional parameter set. 
                Global frequency detuning of $-0.5\ \Gamma$ and a magnetic field gradient of $1$ G/mm are applied. 
                Laser power is equally divided to each frequency component. Polarizations are set as $(+, +, +,-)$.}
        \label{fig:conventional_case}
    \end{figure}

    The maximum capture velocity of the optimal parameter set, identified as $82.5$ m/s, can be contrasted with that of a typical parameter set of a MOT with laser detuning of $0.5\ \Gamma$ to $1\ \Gamma$. 
    Fig.\ref{fig:conventional_case} shows the force and motion trajectories under the conventional parameter set.
    With a capture velocity of $33.75$ m/s, the conventional setup falls significantly short of the optimal parameter set, indicating an improvement of more than twofold in the capture velocity using our method. 
    Molecules exceeding $33.75$ m/s in velocity within the conventional setup tend to accelerate, converging at approximately $60\ \text{m/s}$, and thus escape the MOT. 
    In contrast, molecules with velocities below this capture threshold in the conventional configuration experience scattering forces stronger than those in the optimized cases identified in this study.
    Such observations lead to a potential experimental strategy: employing the optimal parameter set we found here to cool the faster molecules at the beginning, then changing to the conventional parameter set for lowering the temperature further. 
    This approach can be expected to achieve a lower temperature and a stronger restoring force as well as trapping more molecules in a MOT.

    \begin{figure*}[t]
        \centering
        \includegraphics[width=0.8\linewidth]{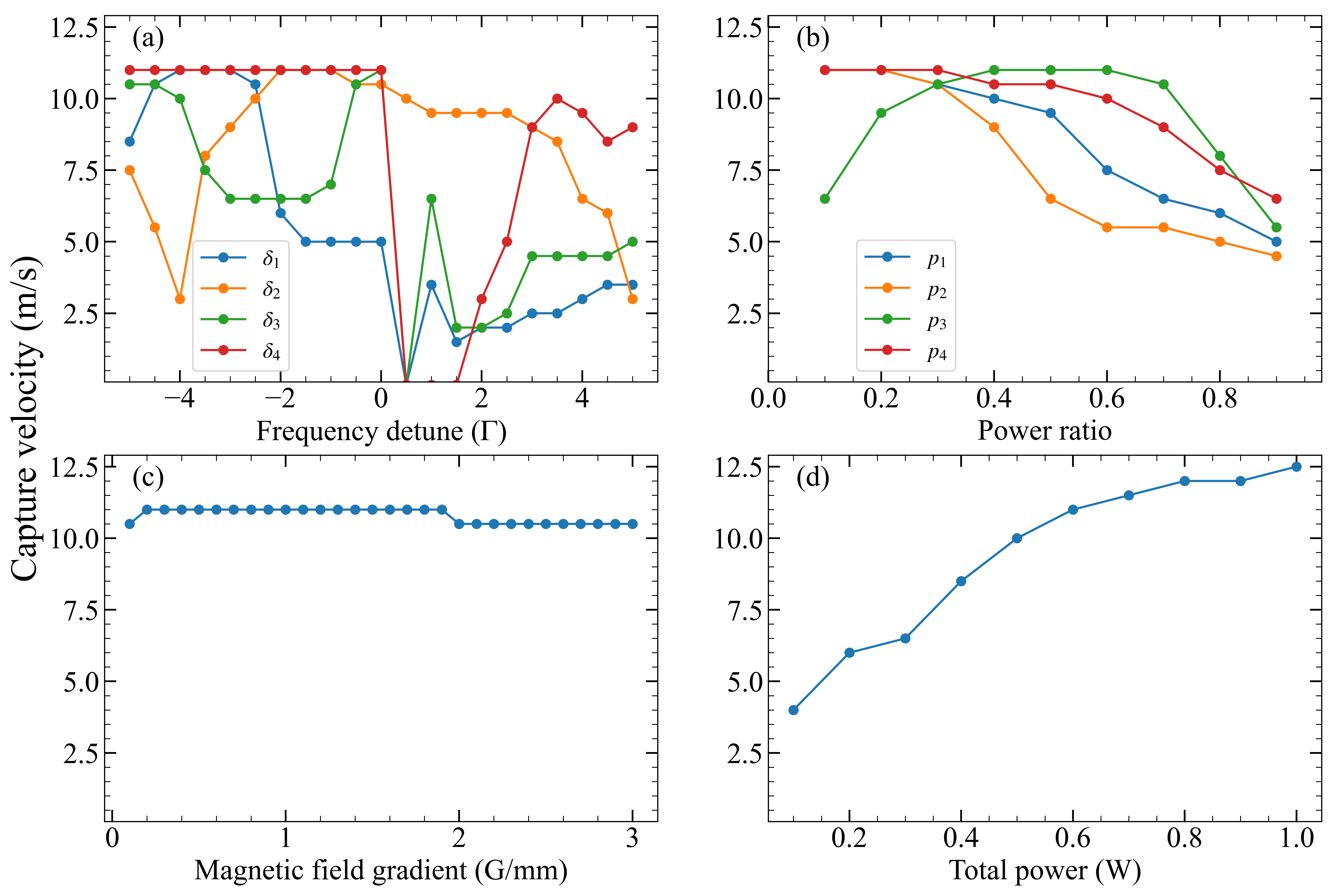}
        \caption{Capture velocity vs change each parameter from $\vec{X}_{\rm mean}$. (a) Frequency detuning, (b) Power ratio of each laser component. (c) Magnetic field gradient. (d) Total power.}
        \label{fig:one_param_variation}
    \end{figure*}

    Fig.\ref{fig:one_param_variation} shows the change of capture velocity $v_c$ when each parameter varies. 
    The frequency detuning of the laser component is shown in Fig.\ref{fig:one_param_variation}(a). 
    For $\delta_1$ and $\delta_2$, the capture velocities are optimal near $\vec{X}_{\rm mean}$ within the $1.5\ \Gamma\simeq30\rm$ MHz range and gradually drop outside the range.
    These ranges are sufficiently broad to be accessible within the resolution limited by natural linewidth $\Gamma$ as well as a linewidth of typical laser systems, thus one can convince that $\delta_1, \delta_2$ are easily adjustable to find optimal point in real experiment.
    However, for $\delta_3$ and $\delta_4$, a sharp dependence is observed near the optimal point.
    $\delta_3$ has only a couple of optimal points near the resonance, and $v_c$ drops to zero when the frequency detuning changes to blue.
    $\delta_4$ also has a sharp drop when the frequency detuning goes to blue, but within the red detuning region $v_c$ is always optimal.
    This implies that both $\delta_3,\delta_4$ should be red detuned, and especially the frequency of laser 3 should be carefully stabilized to get an optimal point.
    Moreover, although the properties of laser 3 and laser 4 are similar because the energy gap between $|F=1^+\rangle$ and $|F=2\rangle$ is indistinguishable, the polarization $\sigma_3$ and $\sigma_4$ can make a difference to the optimal $v_c$.
    For Fig.\ref{fig:one_param_variation} (b), we scan the parameter $p_i$ for the ratio with total power, keeping the power ratio between the other components constant.
    $p_3$ has the largest power ratio more than 40\% during the optimal condition, and other components are lower than 20\%.
    This occurs because laser 3 addresses not only the $|F=1^-\rangle$ state involved in the initial step of the MOT stage along the longitudinal direction (due to the Doppler shift), but also the $|F=1^+\rangle$ and $|F=2\rangle$ states that contribute to the transverse cooling.
    Fig.\ref{fig:one_param_variation} (c) shows the $v_c$ vs magnetic field gradient.
    The change of the magnetic field gradient does not affect $v_c$ from $0.1$ G/mm to $1.9$ G/mm, almost the entire range of scanning with TABLE \ref{table:MOT range}.
    $v_c$ decreases to 10.5 $\Gamma /k$ after 2.0 G/mm, but does not change after this point.
    Fig.\ref{fig:one_param_variation} (d) shows that $v_c$ increases monotonically when the total power increases.

	\subsubsection{Implementing transverse motion of the molecules}\label{subsubsec:implementing transverse motion of the molecules}

    \begin{figure*}[t]
        \centering
        \includegraphics[width=17.0cm]{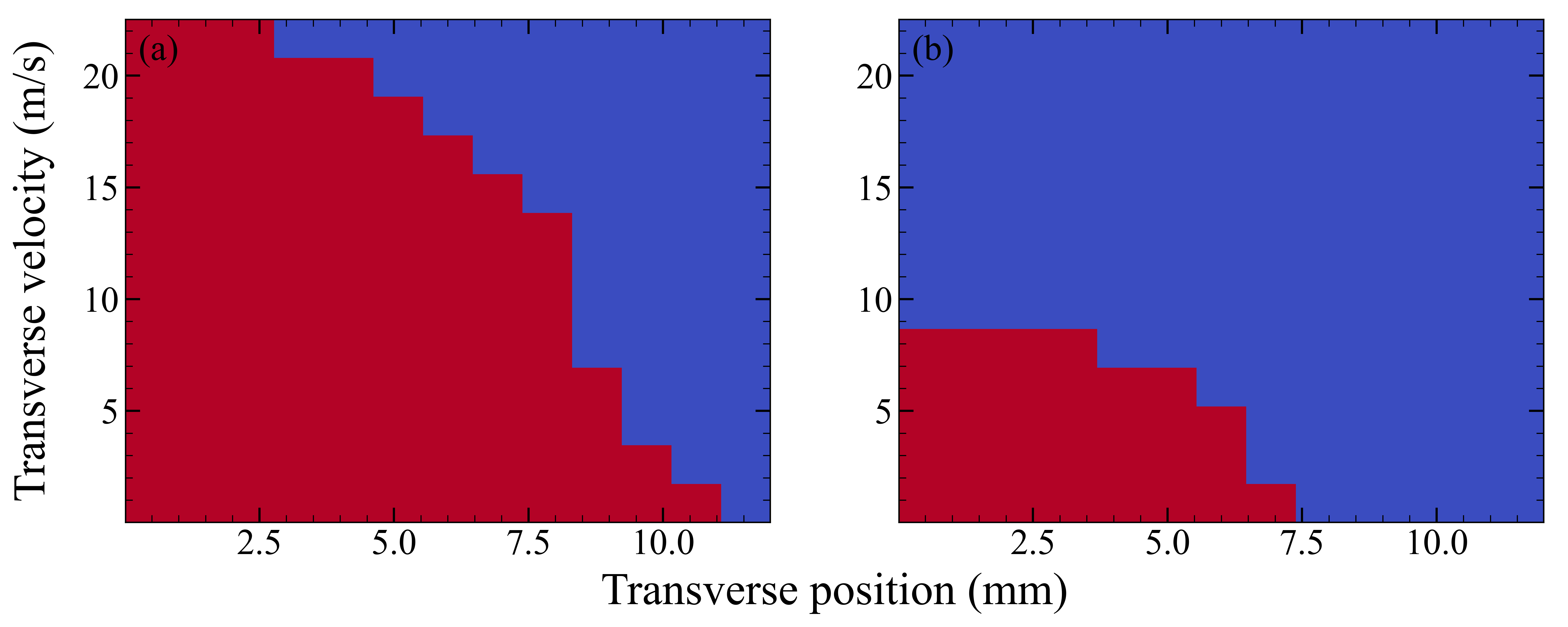}
        \caption{The truth value of the MgF molecule being trapped based on its initial transverse position and velocity with different initial longitudinal velocities of $v_l = 37.5$ m/s (a) and $v_l = 75$ m/s (b). 
        Molecules in the red region are captured in the MOT, while those in the blue region are not.}
        \label{fig:MOT capture condition plot}
    \end{figure*}

    Thus far, all simulations have been based on an ideal case where the transverse motions of the molecules are not taken into account.
    However, whether the molecules are trapped or not strongly depends on the transverse positions and velocities of the molecules.
    To accommodate the transverse conditions into the optimization process, we simulate the motions of the molecules including the transverse positions and velocities in addition to the longitudinal velocities.
    The experimental parameters are set to the mean of the twelve optimal sets, $\vec{X}_{\rm mean}$ that we acquired in the previous step.
    Consistent with the previous process, we consider molecules to be trapped if they are located within 6 mm of the origin along each axis and have velocities below 0.7 m/s toward the center.
    Then we employed interpolation methods to establish a "MOT-trapping condition function" which decides whether the molecules are trapped or not depending on their initial conditions. 
    Fig.\ref{fig:MOT capture condition plot} shows a MOT-trapping region with respect to the transverse positions and velocities of the molecules. 
    The molecules with longitudinal velocity  $v_l = 75$ m/s have a smaller trapping region in the transverse phase space than the molecules with $v_l = 37.5$ m/s.
    This is because the molecules with the higher $v_l$ need the higher laser intensities to be cooled down, so that the molecules should be closer to the longitudinal axis.
    The MOT-trapping condition function gives optimal conditions for the molecules to be trapped with the transverse motion, and this function will be used in the slowing stage to determine whether the molecules are trapped or not.
 
	\subsection{Slowing stage}\label{subsec:slowing stage}

    \subsubsection{Model setup}\label{subsubsec:slowing model setup}
	
    \begin{table}[h]
    \centering
    \caption{Range of parameters preset in the simulation in slowing stage.
    Carrier frequency detuning of the main slowing laser is expressed in units of 20.9 MHz(corresponding to $\Gamma=2\pi\times 20.9$ MHz).
    Modulation indices and frequencies are given in radian and MHz, respectively.
    All time variables are given in ms.}
    \begin{ruledtabular}
    \begin{tabular}{lcc}
        Parameter (unit) & Symbol & Range \\
        \hline
        Carrier frequency detuning  ($\Gamma$) & $\Delta$ & [0, 25] \\
        Modulation index 1 (rad) & $\beta_1$ & [0, 5]  \\
        Modulation index 2 (rad)& $\beta_2$ & [0, 15]  \\
        Modulation frequency 1 (MHz)& $f_1$ & [70, 170]  \\
        Modulation frequency 2 (MHz)& $f_2$ & [5, 44]  \\
        Laser turn-on time (ms)& $t_{\rm on}$ & [1, 3.6]  \\
        Laser operation time (ms)& $t_{\rm operate}$ & [0, 2]  \\
    \end{tabular}
    \end{ruledtabular}
    \label{table:Slowing range}
    \end{table}

    Laser slowing is achieved by two main slowing lasers with opposite polarizations of $\sigma^+$ and $\sigma^-$.
    The two lasers are assumed to be frequency-broadened using two EOMs, one for the hyperfine splittings and the other for white-light generation to cover various Doppler-shifts.
    The slowing lasers are turned on at time $t_{\rm on}$ for a duration of $t_{\rm operate}$, where $t=0$ corresponds to the moment when the molecules are generated.
    The center frequency of the main lasers is red-detuned by a main detuning parameter $\left(\Delta\right)$ to compensate the Doppler shift. 
    Two EOMs are assigned indices $i = 1$(hyperfine EOM) and $2$(white-light EOM), and have different modulation indices $\left(\beta_i \right)$ and modulation frequencies $\left(f_i\right)$. 
    The ranges of parameters are specified in the table \ref{table:Slowing range}.

    The seven parameters, defined as $\vec{\mathbf{X}}_{\rm slow}$, are optimized using Bayesian optimization to capture the maximum number of MgF molecules in a MOT.
    The transverse divergence of the molecular beam should also be considered in the simulation. 
    There are two ways to simulate the motion of the molecules and define a target parameter which should be maximized: 
    (i) \textit{Direct-numerical simulation}; the motion of molecules with all longitudinal and transverse conditions is calculated directly for a given set of parameters. 
    The target parameter is the number of molecules trapped in a MOT.
    Although this method guarantees accuracy, the necessity to continually change the transverse condition during the slowing process makes the process extremely time-consuming. 
    (ii) \textit{Expected-model simulation}; First, we assume that the longitudinal and transverse velocity of our molecular beam has Gaussian distribution, $N(107,74.4)$ m/s for the longitudinal axis and $N(0,74.4)$ m/s for the transverse axis.
    Then we simulate the motional trajectories for the longitudinal axis, by varying the initial longitudinal velocity $v_l$.
    In this case, the expected ratio of molecules captured in a MOT is determined based on the percentage of molecules with $v_l$ to the total velocity distribution and the MOT-trapping condition with final $v_l$.
    The \textit{Expected-model simulation} requires significantly less computational effort compared to the \textit{Direct-numerical simulation}, thereby enabling a more efficient optimization of the slowing stage.
    Thus, we choose the \textit{Expected-model simulation} method.

    To optimize the slowing stage using \textit{Expected-model simulation}, we define a dimensionless ratio, $R_{\rm trap}$, representing the trapping efficiency. $R_{\rm trap}$ is calculated as

    \begin{align}
        R_{\rm trap} &= \dfrac{\rm Ratio\ of\ trapped\ molecules}{\rm Ratio\ of\ molecules\ reaching\ the\ MOT\ region}
    \end{align}

    The numerator, the ratio of trapped molecules can be deduced from the final $v_l$, which represents the longitudinal velocity of the molecules after laser-slowing.
    Using the MOT-trapping condition function defined in the MOT stage simulation, the maximum transverse position and velocity of the molecules to be trapped in a MOT are determined by the final $v_l$.
    The simulation uses two motional trajectories, whose initial transverse position is 0 mm or 12 mm with given initial $v_l$.
    The first case represents the motion where the molecules experience the strongest deceleration, which indicates that the time to reach the MOT stage is the longest. 
    Because the transverse velocity in the slowing stage is constant, the longest time to reach the MOT stage means that the transverse position is the largest and the molecules are least likely to be trapped in a MOT. 
    Under this condition, the least upper bound of the initial transverse velocity where the molecules are trapped in a MOT can be obtained.
    The second case represents the motion where the molecules undergo the least deceleration. 
    Under this condition, the least upper bound of the initial longitudinal velocity where the molecules are trapped in a MOT can be obtained.
    We can derive a conservative lower bound for the ratio between the trapped molecules and the total molecules in a buffer-gas beam.
    
    The denominator of $R_{\rm trap}$ depends on the experimental setup. 
    In our setup, the maximum transverse angle of the molecular beam $\theta_{\rm max} = 6.32^\circ$ is set by the dimensions of the chamber.
    With the initial $v_l$ and $\theta_{\rm max}$, the maximum transverse velocity for the molecules entering the MOT stage $v_{\rm trans}^{\rm max}$ is determined.
    Therefore, the ratio of molecules entering the MOT chamber to the total number of molecules in a buffer-gas beam can be calculated from the velocity distribution of the molecular beam.
    
    Combining these two results, $R_{\rm trap}$ is established as a physical quantity representing the efficiency of the slowing process. 
    Same as the MOT stage, we use the Bayesian optimization method to find the optimal parameter set for the maximum $R_{\rm trap}$.

	\subsubsection{Result}\label{subsubsec:slowing result}
 
    \begin{table*}[t]
        \centering
        \caption{Parameter sets with the best three results from optimization.
        $R_{\rm trap}$ is expressed in percentages.}
        \label{table:slowing result}
        \begin{ruledtabular}
        \begin{tabular}{lcccccccc}
            Case & $\Delta$& $\beta_1$  & $\beta_2$
            & $f_1$ & $f_2$ & $t_{\rm on}$ & $t_{\rm operate}$  & $R_{\rm trap}$\\
            \hline
            1 & 18.82 & 1.173 & 2.490 & 163.29 & 25.24 & 1.567 & 1.238 & 28.6 \\
            2 & 17.92 & 1.234 & 2.689 & 162.76 & 26.71 & 1.671 & 1.008 & 28.5 \\
            3 & 18.87 & 1.294 & 3.788 & 157.08 & 24.82 & 1.520 & 1.069 & 27.9 \\
        \end{tabular}
        \end{ruledtabular}

    \end{table*}

	\begin{figure}[t]
		\centering
		\includegraphics[width=8.5cm]{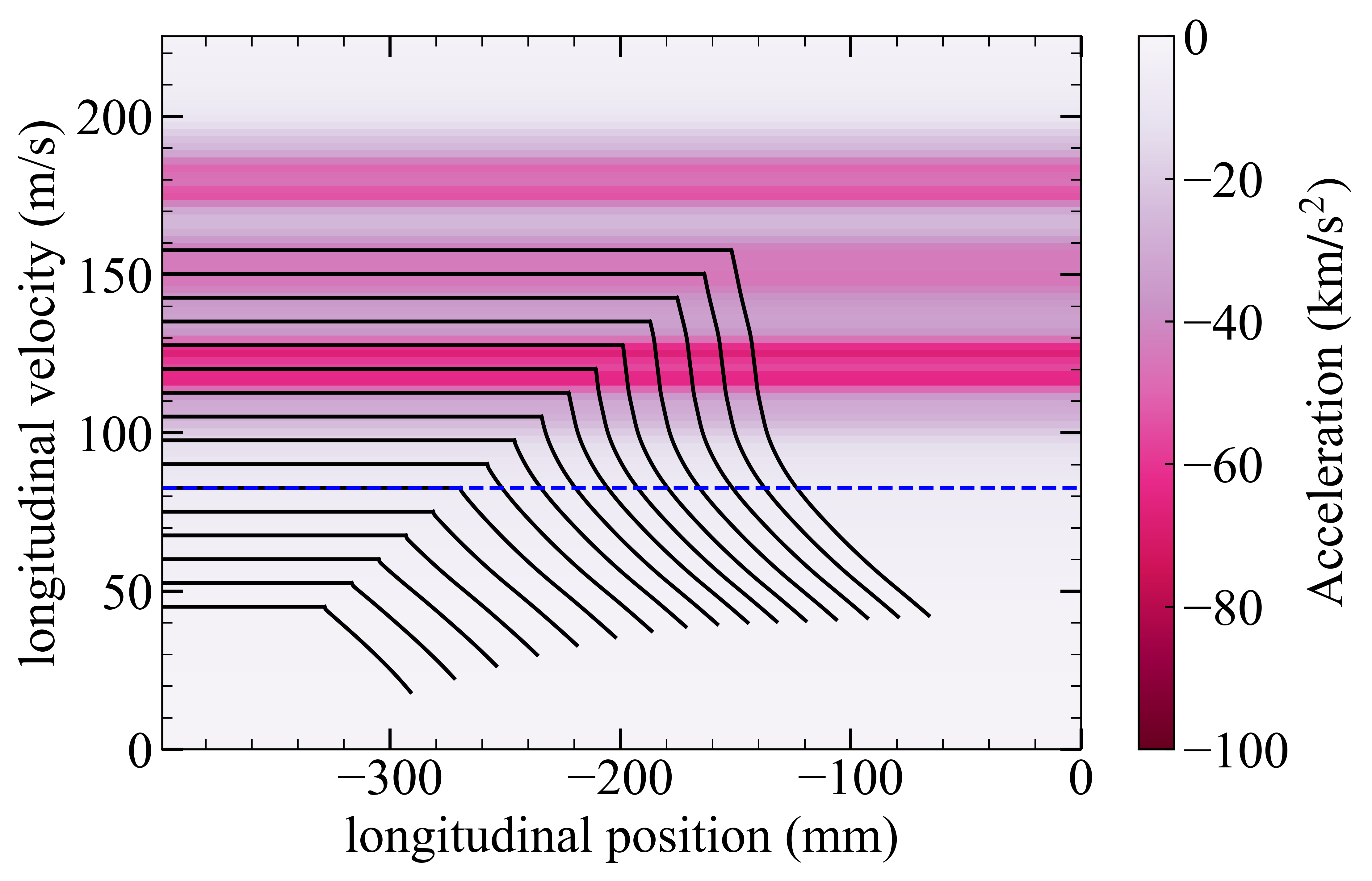}
		\caption{Motional trajectories of the molecules during the slowing stage with the optimal parameters of case 1. Black solid lines show the trajectories of MgF molecules with different initial longitudinal velocities, from $45$ m/s to $157.5$ m/s. MgF molecules moves with constant initial velocity before the slowing lasers are turned on at $t_{\rm on}$. The molecules are decelerated during the time $t_{\rm operate}$ by the slowing lasers. Blue dashed line indicates the capture velocity, $82.5$ m/s determined from MOT stage.}
		\label{fig:Slowing motional trajectories}
	\end{figure}

    Table \ref{table:slowing result} presents the top three results obtained for the slowing stage. 
    The maximum fraction of MgF molecules trapped in a MOT relative to the number of molecules reaching the MOT region is found to be 28.6\%. Fig.\ref{fig:Slowing motional trajectories} illustrates the motional trajectories corresponding to case 1, one of the optimal parameter sets.
    This result shows that the molecules are decelerated below the capture velocity obtained in the MOT stage.

    \begin{figure}[t]
        \centering
        \includegraphics[width=8.5cm]{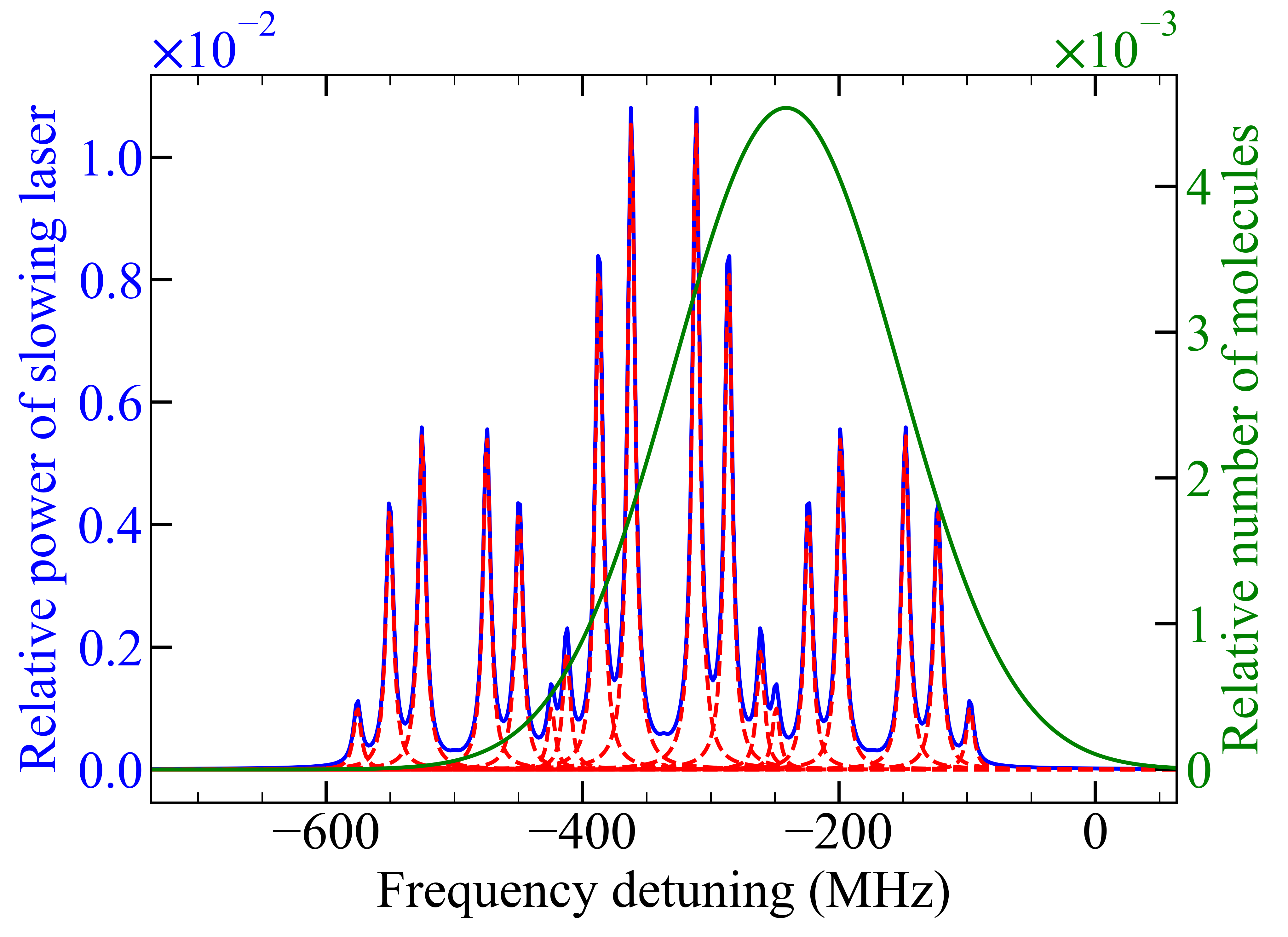   }
        \caption{Spectrum of the slowing laser(blue line) and the resonant-frequency distribution of the initial MgF molecules(green line).
                Origin of the $x$ axis is resonant with frequency from $|F=0\rangle$ of the $X^2\Sigma^+$ state to $A^2\Pi_{1/2}$ state.
                Spectrum of the slowing laser is the sum of the spectrum of each frequency component divided by two EOMs(red line).
                Frequency distribution is calculated from the velocity distribution using the Doppler shift.
                Most of the molecules in $|F=0\rangle$ can be cooled down by the right part of spectrum.
                Different hyperfine states has additional frequency shift from the figure, but the broadened spectrum can also slow down the molecules in different hyperfine states.}
        \label{fig:EOMs spectral broadening}
    \end{figure}

    The spectrum of the slowing laser that is broadened with the optimal parameter set and the distribution of the molecules whose resonant frequencies are shifted due to their initial longitudinal velocities are presented in Fig.\ref{fig:EOMs spectral broadening}.
    Although the peak of the molecular probability distribution is offset from the laser's carrier frequency, the spectral broadening achieved through the two EOMs effectively covers a wide range of molecular Doppler shifts.
    The result in Fig.\ref{fig:EOMs spectral broadening} indicates that most of the laser power is focused on slowing down fast molecules.    
    Relatively slower molecules are less decelerated, allowing for quicker entry into the MOT stage. 

    \begin{figure*}[t]
        \centering
        \includegraphics[width=17.0cm]{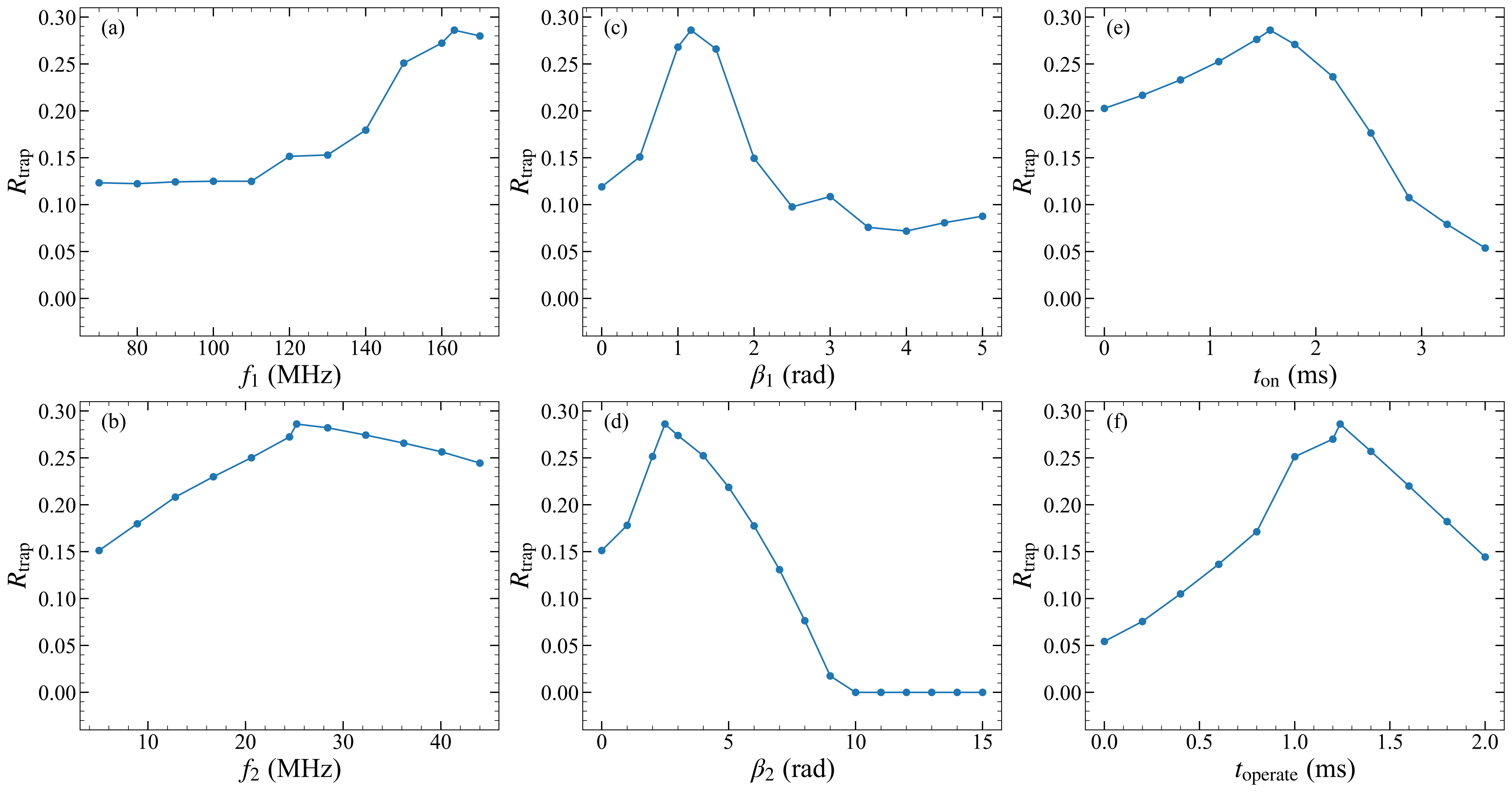}
        \caption{$R_{\rm trap}$ vs each parameter change in slowing stage. (a) and (b) show the effect of varying the modulation frequencies of hyperfine EOM ($f_1$) and the white-light EOM ($f_2$), respectively. Modulation indices are also changed at the both of hyperfine EOM (c) and white-light EOM (d). Laser operating time is also varied for laser-on time (e) and operating time (f).}
        \label{fig:one_param_variation_slowing}
    \end{figure*}

    Fig.\ref{fig:one_param_variation_slowing} shows the dependence of $R_{\rm trap}$ on the variation of each parameter in the slowing stage. 
    In Fig.\ref{fig:one_param_variation_slowing} (a), $f_1$ exhibits a generally increasing trend in $R_{\rm trap}$ upto 150 MHz, indicating that a wider spacing between sidebands enhances the optical cycling.
    Because a white-light EOM can cover a wide range of frequencies, the hyperfine EOM frequencies are optimized by slightly increasing the separation between them compared to the theoretical hyperfine splittings, in order to maximize coverage of the molecular velocity distribution.
    In contrast, $f_2$ shows a clear optimum around $25\sim 30$ MHz, which corresponds to the natural linewidth of the excited state in Fig.\ref{fig:one_param_variation_slowing} (b).
    Around the optimal value of $f_2$, the decrease in $R_{\rm trap}$ is more pronounced when $f_2$ is reduced than increased.
    This asymmetry indicates that the white-light slowing effect becomes more efficient when the modulation frequency of the white-light EOM is larger than the natural linewidth.
    In Fig.\ref{fig:one_param_variation_slowing} (c), the modulation index $\beta_1$ gives the best result at around 1 rad, where the power is efficiently distributed between the zeroth and first order components. 
    Fig.\ref{fig:one_param_variation_slowing}  (d) shows that the optimal range of $\beta_2$ lies between 2 and 3 rad, where most of the power is concentrated in the first and second order components while suppressing the carrier frequency.
    
    In Fig.\ref{fig:one_param_variation_slowing} (c), $R_{\rm trap}$ reaches its maximum when $t_{\rm on} \simeq 1.5$ ms and $t_{\rm operate} \simeq 1.3$ ms, while significant drops are observed outside this region.
    When the laser is turned on earlier (i.e., smaller $t_{\rm on}$), molecules are decelerated at an earlier time, leading to increased transverse spreading due to the extended flight time before reaching the MOT region.
    In contrast, a larger $t_{\rm on}$ results in a faster arrival in the MOT region, but the deceleration duration $t_{\rm operate}$ becomes insufficient, reducing $R_{\rm trap}$. These results emphasize the importance of optimizing the temporal sequence of laser operation in order to achieve the efficient slowing process.
    
    \begin{figure*}[t]
        \centering
        \includegraphics[width=17.0cm]{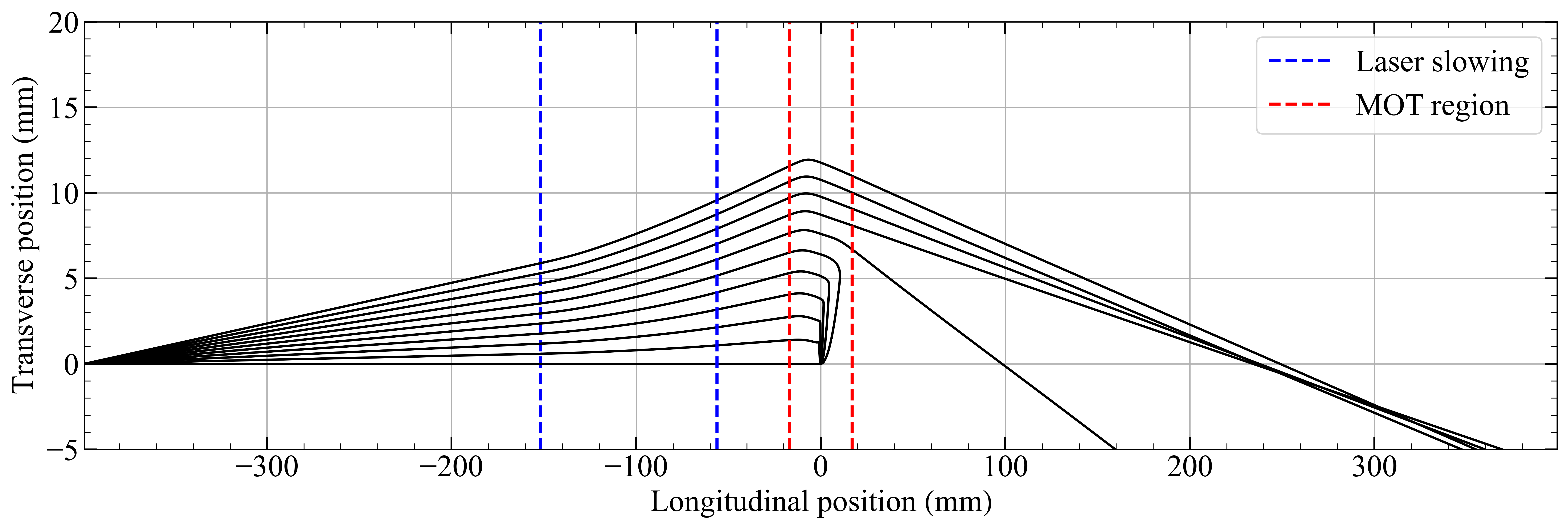}
        \caption{An example of position trajectories using \textit{Direct-numerical simulation} with the optimal parameter set. 
        The Initial longitudinal velocity of molecules is 157.5 m/s, and the transverse velocity varies from 0 to 3.75 m/s.
        Blue dashed lines indicates the region when the slowing lasers are applied, while red dashed line marks the MOT region.}
        \label{fig:total simulation}
    \end{figure*}

    \subsection{Total stage validation}\label{subsec:direct-numerical validation}

    Although \textit{Expected-model simulation} is efficient for the optimization process, \textit{Direct-numerical simulation} is more accurate than \textit{Expected-model simulation}. 
    To verify the validity of our results, we simulate the motion of the total stage with optimized parameters from section \ref{subsec:MOT stage} and \ref{subsec:slowing stage} using the \textit{Direct-numerical simulation}(Fig.\ref{fig:total simulation}).
    This simulation incorporates the full experimental sequence, including both the slowing and MOT stages, and explicitly accounts for random photon recoils arising from spontaneous emission events.
    
    Compared to the \textit{Expected-model simulation}, the trapping efficiency $R_{\rm trap}$ is reduced from 28.6\% to 25.0\%.
    This reduction arises because the random recoil during the slowing stage increases the spread of the transverse momentum, thereby broadening the transverse spatial distribution of the molecules.
    Although the absolute trapping ratio is slightly lower in the \textit{Direct-numerical simulation}, the overall dependence on the optimized parameter sets remains consistent with the \textit{Expected-model simulation}.
    This consistency suggests that the optimization results obtained from the computationally efficient \textit{Expected-model simulation} can be reliably applied under real experimental conditions.

    \section{Conclusion}
    
    In this research, by simulating the MOT and laser slowing process of the MgF molecules altogether, all experimental variables for cooling and trapping the molecules were optimized, and the expected MOT capture velocity and the ratio of trapped molecules were obtained.
    Utilizing the rate equation of PyLCP, we first simulated the motion of moving molecules to determine the conditions for trapping in the MOT. 
    By employing the Bayesian optimization method, we efficiently optimized various experimental variables in the MOT experiment with the expected capture velocity of 82.5 m/s for MgF molecules.
    
    We also optimized the laser slowing process in our simulation.
    Using the specifications of the experimental equipment available in our laboratory, we identified the experimental parameters of the slowing laser that maximized the number of captured molecules in the MOT, demonstrating a $R_{\rm trap}$ of 28.6\%.
    
    However, there are some limitations in our simulation process. 
    In the MOT stage, the simulation only solves the motion equations for a single molecule, thus ignoring factors such as collision between the MgF molecules and the remaining He gas during actual trapping. 
    Additionally, this simulation does not consider vibrational leakage, which causes the molecules to leave the cycle transition.
    In real experiments, how much we repump the vibrational leakage will largely affect the overall optical cycling rate.
    
    Nevertheless, our research provided a set of experimental parameters tailored to MgF molecules to optimize the capture velocity of a MOT and the ratio of captured molecules, as well as insights to understand the trapping and slowing process.
    In the MOT stage, we discovered that the $|F=1^-\rangle$ hyperfine state of the molecule significantly contributes to the cooling process. 
    In the slowing stage, the molecular velocity range to focus for slowing was identified to configure the laser parameters.
    This provides a guide for approaching MOT experiments with MgF molecules.
    Also, by establishing a comprehensive motional trajectory simulation system, we expect to apply these processes to find the necessary conditions for laser cooling experiments with molecules other than MgF.

    \section*{Acknowledgments}

    Special appreciation is given to Kikyeong Kwon, Seunghwan Rho, Youngju Cho, Yongwoong Lee and Hyeonjoon Jang for their fruitful discussion of simulation methods throughout the course of this research. 
    The authors acknowledge support from the National Research Foundation of Korea under grant numbers RS-2022-NR119745, RS-2024-00439981, RS-2024-00431938, and RS-2023-NR068116.

	\bibliography{ref_url}
	
\end{document}